\documentstyle[twoside,fleqn,espcrc2,psfig]{article}
\newcommand{\AmS}{{\protect\the\textfont2
  A\kern-.1667em\lower.5ex\hbox{M}\kern-.125emS}}

\title{$U(1)$ Chiral Gauge Theory with Domain Wall Fermions and Gauge
Fixing}

\author{S. Basak$^{\rm a}$
and
Asit K. De\address{Saha Institute of Nuclear Physics, 1/AF
Salt Lake, Calcutta 700064, India}\thanks{Presenter at Lattice'99.}}

\begin{document}
\begin{abstract}
We investigate a U(1) lattice chiral gauge theory ($L\chi GT$) with the
waveguide formulation of the domain wall fermions and with compact gauge
fixing. In the reduced model limit, there seems to be no mirror chiral modes
at the waveguide boundary.
\end{abstract}
\maketitle
\section{Introduction}
Gauge fixing is not necessary in Wilson lattice gauge theory. For
gauge-non-invariant theories, if gauge fixing is not done with a target
gauge-invariant theory in mind, however, there are nontrivial consequences,
namely, the longitudinal gauge degrees of freedom ({\em dof}) couple to
physical {\em dof}. The well-known example is the Smit-Swift proposal
of $L\chi GT$. The obvious remedy is to gauge fix. The Roma proposal involving
gauge fixing passed perturbative tests but does not address the problem of
gauge fixing of compact gauge fields and the associated problem of lattice
artifact Gribov copies. The formal problem is that for compact gauge-fixing a
BRST-invariant partition function as well as (unnormalized) expectation values
of BRST invariant operators vanish as a consequence of lattice Gribov copies.
Shamir and Golterman \cite{golter1} has proposed to keep the gauge-fixing part
of the action BRST non-invariant and tune counterterms to recover BRST in the
continuum. In their formalism, the continuum limit is to be taken from within
the broken ferromagnetic (FM) phase approaching another broken phase which is
called ferromagnetic directional (FMD) phase, with the mass of the gauge field
vanishing at the FM-FMD transition. This was tried out in a U(1) Smit-Swift
model \cite{bock1} and so far all indications are that in the pure gauge
sector, QED is recovered in the continuum limit and in the reduced model limit
free chiral fermions in the appropriate chiral representation are obtained.

We want to extend this idea to other previous proposals of a $L\chi GT$ which
supposedly failed due to above mentioned coupling of longitudinal gauge
{\em dof} to physical {\em dof}. For this pupose we have chosen the
waveguide formulation of the domain wall fermion where mirror chiral
modes appeared at the waveguide boundary in addition to the chiral modes at
the domain wall or antiwall to spoil the chiral nature of the theory
\cite{golter2}.

\section{Gauge-fixed Domain Wall Action}
For Kaplan's free domain wall fermions on a $4+1$-dimensional lattice of
size $L^4 L_s$ where $L_s$ is the 5th dimension, with periodic boundary
conditions in the 5th or $s$-direction, there is always an anti-domain wall
and the model possesses two zero modes of opposite chirality, one bound to
the domain wall at $s=0$ and the other bound to the antiwall at
$s=\frac{L_s}{2}$. With the domain wall mass $m(s)$ taken as
\[ m(s) = \begin{array}{cl}
-m_0 & ~~~~~0 <s< \frac{L_s}{2} \\
 0   & ~~~~~s = 0, \frac{L_s}{2} \\
 m_0 & ~~~~~\frac{L_s}{2} <s< L_s
\end{array} \]
the wall mode is lefthanded (LH), the antiwall mode is righthanded (RH). If
$m_0 L_s\gg 1$, these modes would have exponentially small overlap. These
chiral modes exist for momenta $p$
below a critical momentum $p_c$, i.e. $\vert \hat{p} \vert < p_c$, where
$\hat{p}^2 = 2\sum_\mu[1-cos(p_\mu)]$ and $p_c^2 = 4 - 2m_0/r$.

Since the LH and RH modes of the fermion are separated in
$s$ space, one can attempt to couple these two in different ways
to a gauge field. A 4-dimensional gauge field which is same for all
$s$-slices can be coupled to fermions only for a restricted number of
$s$-slices around the anti-domain wall \cite{golter2} with a view to coupling
only to the RH mode at the antiwall. The gauge field is thus confined
within a {\em waveguide}, $WG = (s: s_0 < s \leq s_1)$.
Our choice of $s_0$ and $s_1$ is such that the anti-domain wall is
located at the center of the $WG$, and at the same time far enough from
the domain or anti-domain wall that the zeromodes are exponentially small at
the $WG$ boundary \cite{golter2}: $s_0 = \frac{L_s+2}{4}-1$,
$s_1 = \frac{3L_s+2}{4}-1$. With this choice $L_s-2$
has to be a multiple of four. The global symmetry of the model and the gauge
transformations on the fermion field in and outside $WG$ remain exactly
the same as in ref. \cite{golter2}.
Obviously, the hopping terms from $s_0$ to $s_0+1$ and that from $s_1$ to
$s_1+1$ would break the local gauge invariance of the action. This
is taken care of by gauge transforming the action and thereby picking up the
pure gauge {\em dof} or a radially frozen scalar field $\phi$
(St\"{u}ckelberg field) at the $WG$ boundary. So
the gauge invariant $WG$ action now becomes, retaining only the $s$
index, taking the Wilson $r=1$ and lattice constant $a=1$,
\begin{eqnarray} S_{F}\!\!\!\!\! & = & \!\!\!\!\!\!\! \sum_{s \in WG}
\overline{\psi}^s \left( D\!\!\!\!/~(U) - W(U) + m(s) \right) \psi^s \nonumber
\\ &+& \!\!\!\!\! \sum_{s\not\in WG} \overline{\psi}^s \left( \partial\!\!\!/
- w + m(s) \right) \psi^s \nonumber \\ &-& \!\!\!\!\!\!\! \sum_{s\neq s_0,s_1}
\left( \overline{\psi}^s P_L \psi^{s+1} + \overline{\psi}^{s+1} P_R \psi^s
\right) + \sum_s \overline{\psi}^s \psi^s \nonumber \\
&- & \!\!\! y \left( \overline{\psi}^{s_0} \phi^\dagger P_L \psi^{s_0+1} +
\overline{\psi}^{s_0+1} \phi P_R \psi^{s_0} \right) \nonumber \\
& -& \!\!\! y \left( \overline{\psi}^{s_1} \phi P_L
\psi^{s_1+1} + \overline{\psi}^{s_1+1} \phi^\dagger P_R \psi^{s_1} \right)
\label{wgact} \end{eqnarray}
where $y$ is the Yukawa coupling at the $WG$ boundaries and the projector
$P_L$ ($P_R$) is $(1-\gamma_5)/2$ ($(1+\gamma_5)/2$). The $D\!\!\!\!/~(U)$ and
$W(U)$ are the gauge covariant Dirac operator and the Wilson term
respectively.

The gauge-fixed pure gauge action for U(1), where the ghosts are free and
decoupled, is:
\begin{equation}
S_B(U) = S_g(U) + S_{gf}(U) + S_{ct}(U) \label{ggact}
\end{equation}
where, $S_g$ is the usual Wilson plaquette action,
gauge fixing term $S_{gf}$ and the gauge field mass counter term
$S_{ct}$ are given by,
\begin{eqnarray}
S_{gf}(U)\!\!\!\! & = & \!\!\!\!\tilde{\kappa} \left( \sum_{xyz} \Box(U)_{xy}
\Box (U)_{yz} - \sum_x B_x^2 \right), \label{gfact} \\
S_{ct}(U) \!\!\!\! & = & \!\!\!\! - \kappa \sum_{x\mu} \left( U_{\mu x} +
U_{\mu x}^\dagger \right),
\end{eqnarray}
\begin{figure}[t]
\begin{center}
\psfig{figure=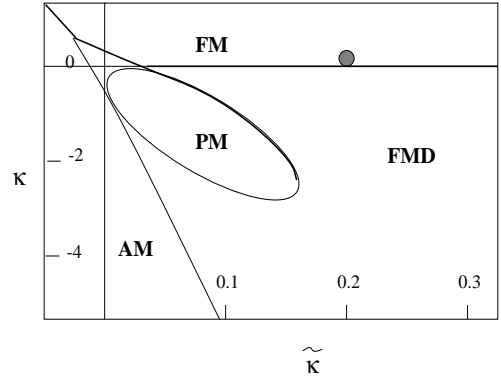,width=6.5cm,height=5.0cm}
\vspace{-1.0cm}
\caption{{\em Schematic quenched phase diagram}}
\end{center}
\vspace{-1.2cm}
\end{figure} 
where $\Box(U)$ is the covariant lattice laplacian and
\begin{equation}
B_x  = \sum_\mu \left( \frac{V_{\mu x-\hat{\mu}} +
V_{\mu x}}{2} \right)^2  \label{bxsq}
\end{equation}
with
$V_{\mu x} = \frac{1}{2i} \left( U_{\mu x} - U_{\mu x}^\dagger \right)$
and $\tilde{\kappa} = 1/(2\xi g^2)$.
$S_{gf}$ has a unique absolute minimum at $U_{\mu x}=1$, validating
weak coupling perturbation theory (WCPT) around $g=0$ or
$\tilde{\kappa}=\infty$ and in the naive continuum limit it reduces to
$\frac{1}{2\xi} \int d^4x (\partial_\mu A_\mu)^2$.

Obviously, the action $S_B(U)$ is not gauge invariant. By giving it a gauge
transformation the resulting action $S_B(\phi^\dagger_x U_{\mu x}
\phi_{x+\hat{\mu}})$ is gauge-invariant with $U_{\mu x}\rightarrow g_x U_{\mu
x} h^\dagger_{x+\hat{\mu}}$ and $\phi_x \rightarrow g_x \phi_x$, $g_x \in
U(1)$. By restricting to the trivial orbit, we arrive at the so-called
{\bf reduced model} action
\begin{equation}
S_{reduced} = S_F(U=1) + S_B(\phi^\dagger_x \;1 \;\;\phi_{x+\hat{\mu}})
\label{reduced}
\end{equation}
where $S_B(\phi^\dagger_x \;1 \;\;\phi_{x+\hat{\mu}})$ now is a
higher-derivative scalar field theory action.

\section{Results in the Reduced Model}
In the quenched approximation, we have first numerically confirmed the
{\mbox phase diagram} in \cite{bock1} of the model in
($\kappa,\tilde{\kappa}$) plane. The phase diagram shown schematically in
Fig.1 has the interesting feature that for large enough $\tilde{\kappa}$,
there is a continuous phase transition from FM to FMD phase. FMD phase is
characterised by loss of rotational invariance and the continuum limit is to
be taken from the FM side of the transition. In the full theory with dynamical
gauge fields, the gauge symmetry reappears at this transition and the gauge
boson mass vanishes, but the longitudinal gauge {\em dof} remain decoupled.
\begin{figure}[t]
\begin{center}
\vspace{-1.8cm}
\hspace*{-0.9cm}\psfig{figure=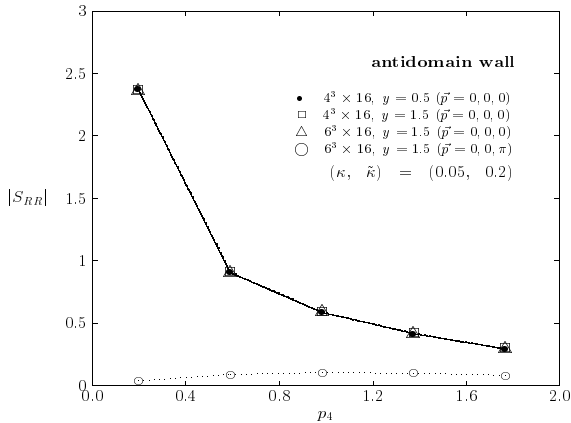,width=8.5cm,height=10.5cm}
\vspace{-4.7cm}
\caption{{\em $RR$-propagator at  antidomain wall ($L_s=22$; a.p.b.c. in
$L_4$)}}
\end{center}
\vspace{-1.31cm}
\end{figure} 
For calculating
the fermion propagators, we have chosen the point $\kappa =0.05$,
$\tilde{\kappa}=0.2$ (gray blob in Fig.1). Numerically  on $4^3 16$ and $6^3 16$
lattices with $L_s=22$ we look for chiral modes at the domain wall ($s=0$),
the antidomain wall ($s=11$), and at the $WG$ boundaries ($s=5,6$ and
$s=16,17$). Error bars in Figs.2 and 3 are smaller than the symbols. Fig.2
shows the $RR$-propagator $|S_{RR}|$ at $s=11$ as a function of a component of
momentum $p_4$ for both $\vec{p}=(0,0,0)$ (physical mode) and $(0,0,\pi)$
(first doubler mode) at different $y$-couplings. From the figure, it is clear
that the doubler does not exist and the physical $RR$-propagator seems to have
a pole at $p=(0,0,0,0)$. The curve stays the same irrespective of
$y$-coupling and lattice size. We have also checked that it coincides exactly
with the free $RR$-propagator (corresponding to $y\phi =1$) at $s=11$. Also
similar analysis with the $LL$-propagator at $s=11$ does not show
any pole. We conclude that there is {\em only a free RH fermion} at the
antidomain wall. From similar data not shown here, we conclude that at the
domain wall, there is a {\em only free LH fermion}. \begin{figure}[t]
\begin{center}
\vspace{-1.8cm}
\hspace*{-0.9cm}\psfig{figure=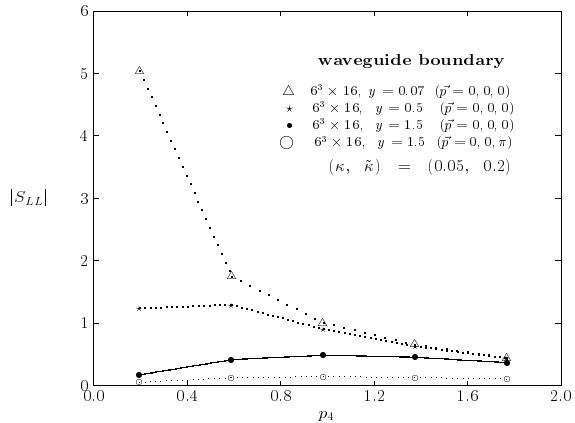,width=8.5cm,height=10.5cm}
\vspace{-4.7cm}
\caption{{\em $LL$-propagator at waveguide boundary ($L_s=22$, a.p.b.c. in
$L_4$)}}
\end{center}
\vspace{-1.2cm}
\end{figure}
Fig.3 shows the $LL$ propagator at $s=6$ (waveguide boundary, just inside).
Here too doublers do not exist. While there is a hint of a pole for $y=0.07$,
the data for $y=0.5,~1.5$ does not favor any pole at zero momentum. Taking
all our numerical data into account for the boundary propagators, there does
not seem to exist any chiral modes there for $y\sim 1$. For very small $y$
the situation is tricky, because strictly at $y=0$, there are mirror boundary
modes present, as can be seen from considerations of fermion current in
$s$-direction and also from numerical simulation. We have done WCPT around
$\tilde{\kappa}=\infty$ and $y=1$ and it supports the numerical data for
$y\sim 1$.

In spite of having the longitudinal gauge {\em dof} explicitly in the action,
it seems that in the reduced model we end up only with free undoubled chiral
fermions at the domain wall and the antiwall with no mirror modes at the
$WG$ boundaries.

\end{document}